\documentclass[referee]{raa}            

\usepackage{graphicx,times}             
\usepackage{natbib}
\usepackage{amssymb,amsmath}
\usepackage{tablefootnote}
\bibpunct{(}{)}{;}{a}{}{,}

\usepackage[pagebackref=true]{hyperref}
\hypersetup{colorlinks = true, linkcolor = green, anchorcolor = red, citecolor = blue, filecolor = red, pagecolor = red, urlcolor = red}

\begin{document}

   \title{Searching for pulsars with phase characteristics}

   \volnopage{Vol.0 (20xx) No.0, 000--000}      
   \setcounter{page}{1}          

   \author{
     Bo Peng$^{\ast}$\inst{1}
  \and Qian-Chen Hu\inst{1}
  \and Qiang Li\inst{1}
   \and Lei Qian\inst{2,3,4,5}
   \and Xiao-Bo Dong\inst{6}
    \and Shi-Lin Peng\inst{1}
    \and Ze-Lin Wang\inst{1}
   }

   \institute{School of Information Engineering, Southwest University of Science and Technology, Mianyang 621010, China; {\it pengball@swust.edu.cn}
    \and
    National Astronomical Observatories, Chinese Academy of Sciences, Beijing 100101, China
    \and 
    CAS Key Laboratory of FAST, National Astronomical Observatories, Chinese Academy of Sciences, Beijing 100101, China
    \and
    State Key Laboratory of Space Weather, Chinese Academy of Sciences,
    Beijing 100190, China
    \and
    University of Chinese Academy of Sciences, Beijing 100049, China
    \and
    Yunnan Observatories, Chinese Academy of Sciences, Kunming, Yunnan 650011, China
\vs\no
   {\small}}

\abstract{We present a method by using the phase characteristics of radio observation data for pulsar search and candidate identification. The phase characteristics are relations between the pulsar signal and the phase correction in the frequency-domain, and we regard it as a new search diagnostic characteristic. Based on the phase characteristics, a search method is presented: calculating DM (dispersion measure) -- frequency data to select candidate frequencies, and then confirming of candidates by using the broadband characteristics of pulsar signals. Based on this method, we performed a search test on short observation data of M15 and M71, which were observed by Five-hundred-meter Aperture spherical radio Telescope (FAST), and some of the Galactic Plane Pulsar Snapshot survey (GPPS) data. Results show that it can get similar search results to PRESTO (PulsaR Exploration and Search TOolkit) while having a faster processing speed. 
}

   \authorrunning{Bo Peng et al.}           
   \titlerunning{Searching for pulsars with phase characteristics} 
   \maketitle

\section{Introduction}
\label{sect:intro}
Pulsars are important in the studies of space navigation (e.g., \citealt{buist2011overview}), gravitational wave detection (e.g., \citealt{lommen2015pulsar}), and interstellar medium (e.g., \citealt{manchester1977pulsars}). For these works, the detection and discovery of new pulsars is meaningful. To a typical pulsar search process, it consists of several steps, i.e., removing radio frequency interferences (RFIs), dedispersion, periodic signal and single pulse searches, and candidate identification. 
In these processing steps, a large number of intermediate data with different dispersion measures (DMs) will be generated. 
On the generated data, a further search with different algorithms will be carried out, such as the fast folding algorithm (FFA; \citealt{staelin1969fast}) for long-period signals, acceleration search (\citealt{ransom2000binary}) and phase modulation algorithm (\citealt{jouteux2002searching}) for binary pulsars.
With such a large amount of intermediate data and complex search processing, it is a time-consuming process to get those pulsar candidates. As an example, it takes 170 second to process a data set of 26 second observation with a 24-cores computing (\citealt{yu2020presto}). To this end, a bunch of accelerated schemes have been provided, but most of them are carried out under the existing framework, such as using GPU to accelerate on the dedispersion (\citealt{barsdell2012accelerating}) and acceleration search (e.g., PRESTO\footnote{\url{http://www.cv.nrao.edu/~sransom/presto/}} (PulsaR Exploration and Search TOolkit) on gpu\footnote{\url{https://github.com/jintaoluo/presto_on_gpu}} by Luo et al. in prep), using FPGA as a backend for pulsar dedispersion processing (\citealt{luo2017digital}), or using computer clusters to parallelize the entire search (\citealt{yu2020presto}). These methods have a relatively significant improvement in computing speed, but there are  some problems in their application, such as how to deal with the memory computing bottleneck (\citealt{sclocco2016real}) when using a single machine for dedispersion. 

Some new algorithms like Fourier-domain dedispersion (FDD; \citealt{bassa2022fourier}), could improve the efficiency of data processing and avoid computational bottlenecks at the same time. But, these new algorithms mainly improve parts of the processing structure rather than the whole search framework. In that way, the overall speed can not be fully improved when processing the bountiful data emerged in the search. Inspired by FDD, a phase characteristic can be calculated with the discarded phase data after FDD calculation. And a DM-frequency image is obtained, showing the influence of DM change on different frequencies. On the basis of FDD, DM-frequency data of the entire data space can be obtained by adding a small amount of calculation, and a method for candidate frequency screening is introduced. With the increase of the observation channels of modern telescopes (\citealt{lyon2016fifty}), the phase characteristic can clearly show the situation of the signals in different channels and the broadband and dispersion features of the pulsar signals can also be reflected. According to those features or characteristics, the pulsar candidate signals can be judged and confirmed. Combined with FDD, phase characteristic can also be integrated into the traditional search structure, by which broadband characteristic images can be quickly obtained. In order to test, the data of M15, M71 and some of GPPS (\citealt{han2021fast}) data observed by FAST(\citealt{nan2006science}; \citealt{jiang2019commissioning}; \citealt{jiang2020fundamental}; \citealt{qian2020fast}) is used for phase characteristic calculation.

In this paper, we introduced the detailed description of phase characteristic in Section \ref{sect:PHASE CHARACTERISTIC}. Section \ref{sect: results} presented the search test. The discussion and conclusion are given in Section \ref{sect:conclusion}.

\section{METHODS}
\label{sect:PHASE CHARACTERISTIC}
\subsection{Related Work}
\label{sect:related work}
A typical time-domain dedispersion algorithm can be described as shifting raw data (a 2D array of fast sampled time series across multiple frequency channels) with dispersion delay based on DM value, and summing up all channels' data. Equation \eqref{eq:time delay} is the dispersion delay between frequency $v_1$ and $v_2$ at a specific DM, based on the Tools of Radio Astronomy (\citealt{rohlfs2013tools}) . 

\begin{equation}
    \frac{d}{\mu s} = 4.148\times 10 ^{9}  
    \begin{bmatrix}
    	\frac{DM}{cm^{-3}pc}
    \end{bmatrix}
    \begin{bmatrix}
    \frac{1}{(\frac{v_1}{MHz})^2}- \frac{1}{(\frac{v_2}{MHz})^2}
    \end{bmatrix}
    \label{eq:time delay}
\end{equation}

Generally, assuming $N$ is the number of channels, channel 1 is the standard phase, and $d_n$ is the dispersion delay under the specific DM between channel 1 and channel n. Then, time-domain dedispersion can be described by Equation \eqref{eq:time-domain dedispersion}, in which $S_c$ is dedispersed data, and $S_n$ is the raw data of channel n. Obtaining $S_c$ is a low computational density process, which only requires a few computation time, while most time is used for raw data reading from memory. As the DM is unknown in the process of pulsar search, raw data will be read many times to traverse all possible DMs.

\begin{equation}
	S_c(t) = \sum_{n = 1}^{N} S_n(t-d_n)
	\label{eq:time-domain dedispersion}
\end{equation}

Based on the time-domain dedispersion algorithm, FDD chooses a different way of data processing. 
During FDD, Fast Fourier Transform (FFT) is directly performed on the raw data of each channel as Equation \eqref{eq:fdd}, in which $X_n(\omega)$ is the FFT result of $S_n(t)$, and $\omega = \{ \omega _i | \omega _i = \frac{2\pi i}{T}, i\in [1, \frac{T}{2}] \}$, $T$ is the number of sampling points. As $\omega _i$ is determined by point index $i$ in frequency domain, we simply use frequency $\omega _i$ refer to the frequency corresponding to the point with index $i$.
Then, based on the equivalence of frequency-domain phase shift and time-domain shift (\citealt{bracewell1986fourier}), rotation factor is used to achieve the effect of dedispersion (see Equation \eqref{eq:fdd2}). With the correction of rotation factor $e^{j\omega d_n}$, frequency-domain dedispersed data $X_c(\omega)$ can be obtained. 
At the end of FDD, $X_c(\omega)$ will be turned to time domain dedispersed data $S_c$ with Inverse Fourier transform.

\begin{equation}
   S_{raw} = \begin{bmatrix}
        S_1(t) \\
        S_2(t) \\
        \vdots \\
        S_N(t) \\ 
    \end{bmatrix} 
     \xrightarrow{\mathcal{F}}
    \begin{bmatrix}
   			 X_{1}(\omega) \\
    		 X_{2}(\omega) \\
   			 \vdots \\
   			 X_{N}(\omega) 
    \end{bmatrix}
    \label{eq:fdd}
\end{equation}

\begin{equation}
    \begin{bmatrix}
    	X_{1}^\top(\omega)&
        X_{2}^\top(\omega)&
      	\cdots&
      	X_{N}^\top(\omega) 
    \end{bmatrix}
    \times
    \begin{bmatrix}
        e^{j\omega d_1} \\
        e^{j\omega d_2} \\
  		\vdots          \\
    	e^{j\omega d_N}
    \end{bmatrix}=X_c(\omega)
    \xrightarrow{\mathcal{F}^{-1}} S_c(t)
    \label{eq:fdd2}
\end{equation}

\begin{figure}[htb]
   \centering
   \includegraphics[width=10cm, angle=0]{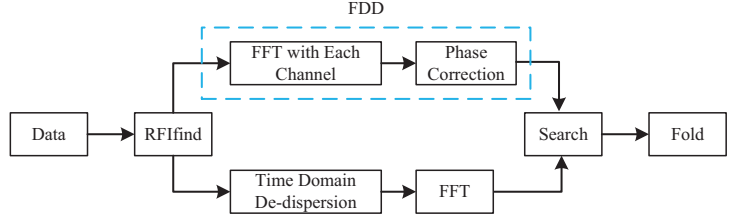}
   \caption{Search process with FDD and time-domain dedispersion.}

   \label{fig:search process}
\end{figure}

Although the processing results of FDD are still time-domain, the frequency-domain dedispersed data $X_c(\omega)$ is obtained during the process, which can be directly used in pulsar search, as shown in Figure \ref{fig:search process}. Compared with the traditional way, FDD has obtained frequency-domain data $X_n(\omega)$ of each channel, which contains a lot of phase information. However, the phase information of $X_n(\omega)$ is only used for dedispersion in FDD. 

\newpage

\subsection{Phase Alignment}
\label{sect:Phase Alignment}
Theoretically, without considering the influence of noise and radio frequency interferences (RFIs), if there is only one pulsar with frequency $\omega _i$ in the observation data, there will be a peak at frequency $\omega _i$ in the power spectrum of each observation channel, that is, $|X_n(\omega _i)| \ge |X_n(\omega _m)|, m \neq i$. 
However, due to the dispersion, if we directly superimpose these complex vectors like $\sum_{n=1}^N{X_n(\omega _i)}$, the peak will no longer be significant.
Considering the influence of dispersion as shown in Equation \eqref{eq:fdd2}, the complex vector sum at frequency $\omega _i$ is turned to $X_c(\omega _i) = \sum_{n=1}^N{X_n(\omega _i)e^{j\omega _i d_n}}$, in which the phase angle of complex vector $X_n(\omega _i)$ is corrected with $e^{j\omega _i d_n}$.
Note that $d_n$ is determined by DM, this means in time-domain dedispersion, correct DM aligns the pulses of each channel, while in frequency-domain dedispersion, correct DM aligns complex vectors of each channel at pulsar frequency $\omega _i$.

\begin{figure}[htb]
	\begin{minipage}[t]{0.32\linewidth}
		\centering
		\includegraphics[width=48mm]{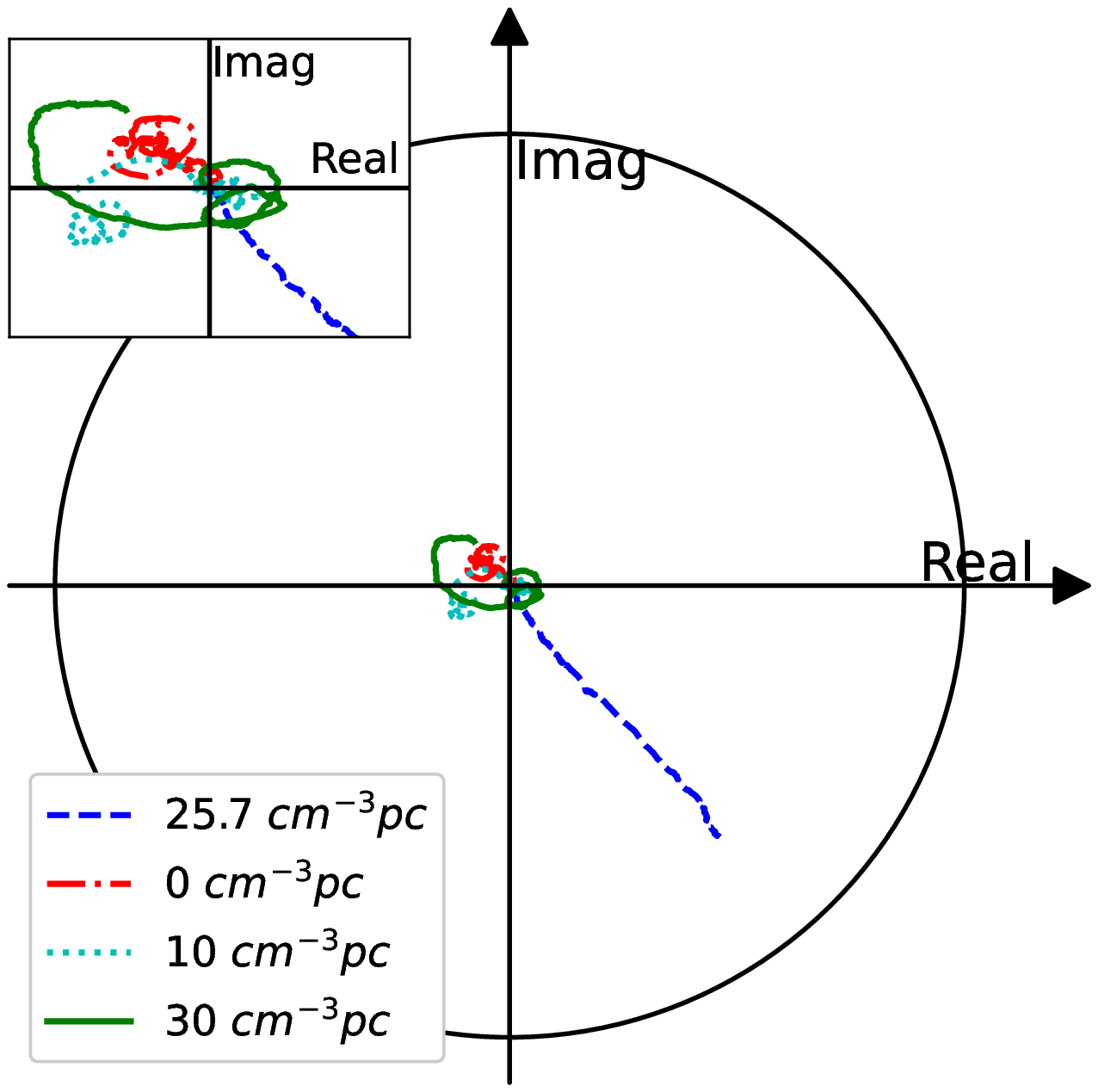}
		\centerline{(a) PSR J1905+0400, 264.2423 Hz}
	\end{minipage}
	\begin{minipage}[t]{0.32\textwidth}
		\centering
		\includegraphics[width=48mm]{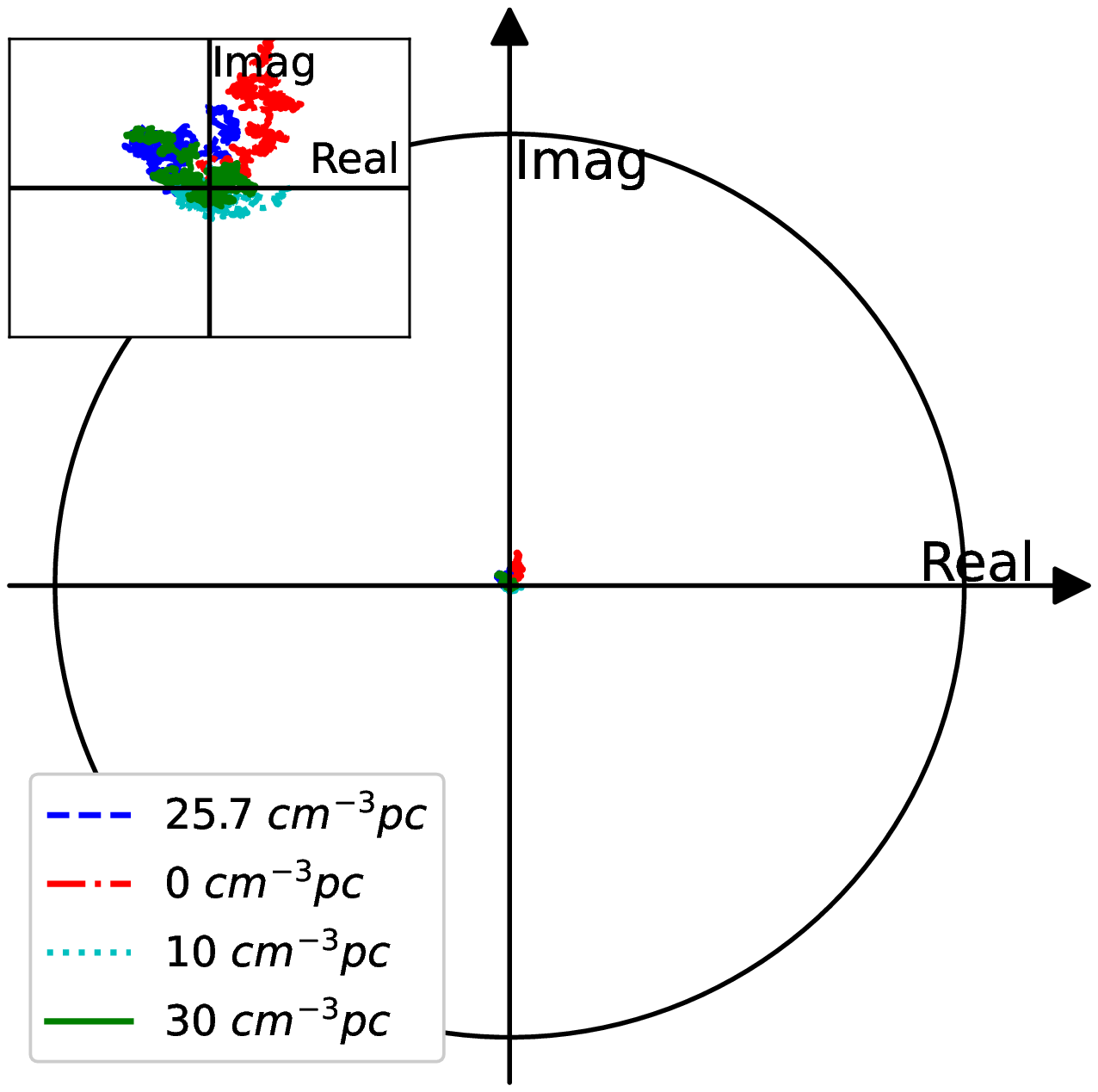}
		\centerline{(b) 200 Hz}
	\end{minipage}
	\begin{minipage}[t]{0.32\textwidth}
		\centering
		\includegraphics[width=48mm]{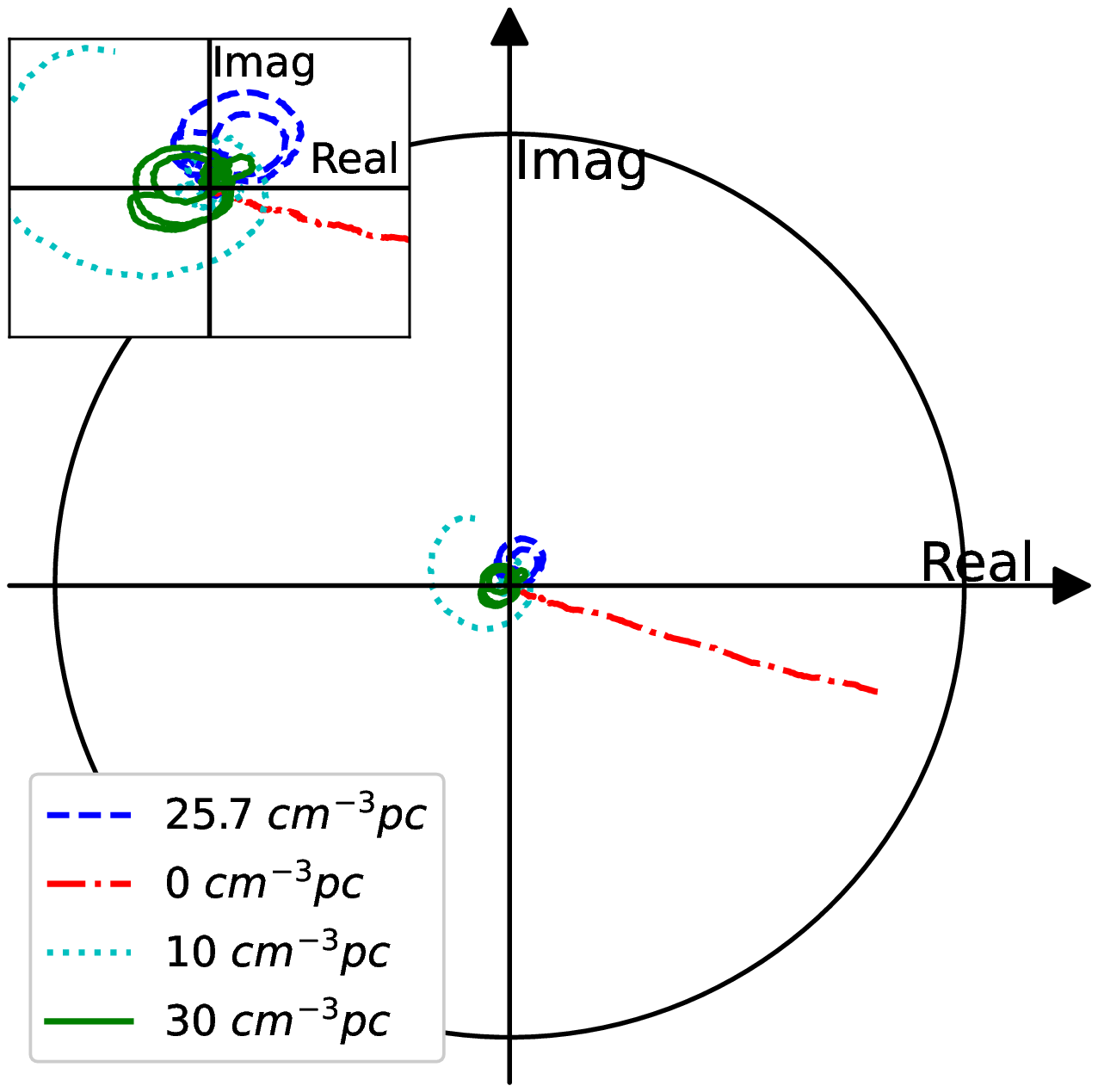}
		\centerline{(c) 150.1046 Hz}
	\end{minipage}
	\caption{Complex vector sum at $\rm264.2423 Hz$, $\rm200 Hz$ and $\rm150.1046 Hz$ with different DMs.
	 $\rm200 Hz$ is chosen as the noise frequency, limited by the number of samples, the actual frequency is 199.9982Hz, which is very close to the terrestrial mains utility frequency's harmonic. $\rm150.1046 Hz$ is terrestrial mains utility frequency's third harmonic.
		The outer black circle is $\sum_{n=1}^N{|X_n(\omega _i)|}$, in which $\omega _i$ is $\rm264.2423 Hz$ in (a), $\rm200 Hz$ in (b) and $\rm150.1046 Hz$ in (c), and curves in (a), (b) and (c) show the superposing process of complex vectors of each channel corrected with different DMs.
		At $\rm264.2423 Hz$, the frequency of PSR J1905+0400, correct DM $\rm25.7 cm^{-3} pc$ makes the vector sum align and closest to the outer black circle.}
	\label{fig:J1905+0400 phase align}
\end{figure}

That alignment effect can be shown in Figure \ref{fig:J1905+0400 phase align}, which is drawn with one of the GPPS data (\cite{ransom2002fourier} has drawn a similar image to prove the effect of frequency derivative correction). 
Figure \ref{fig:J1905+0400 phase align} shows that at pulsar frequency 264.2423Hz, the correct DM aligns complex vectors of each channel, making the length of the complex vector sum to be the longest (see the dashed line in Figure \ref{fig:J1905+0400 phase align}(a)). 
However, when using non-pulsar frequency or uncorrected DM, the trajectory of complex vector sum is around the center (see the other line in Figure \ref{fig:J1905+0400 phase align}). Compared with the complex vector sum with non-pulsar frequency or uncorrected DM, the complex vector sum at $\rm264.2423 Hz$ and $\rm25.7 cm^{-3} pc$ has shown remarkable feature of phase alignment, which could be used as a basic component of the pulsar phase characteristics.

\subsection{Phase Characteristic for Search}
\label{sect:Phase Characteristic for Search}
\begin{figure}[htb]
	\centering
	\includegraphics[width=7.5cm, angle=0]{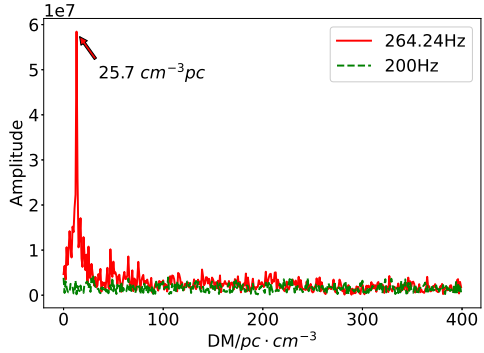}
	\caption{Use 0-1000 $\rm {cm}^{-3}\rm{pc}$ as DM with a step size of 1 $\rm {cm}^{-3}\rm{pc}$ to correct the signals in  Figure \ref{fig:J1905+0400 phase align}, then superimpose the corrected data of each channel and take the modulus as amplitudes.}
	\label{fig:J1905+0400 phase correct}
\end{figure}

Since Figure \ref{fig:J1905+0400 phase align} contains only a few DMs, Figure \ref{fig:J1905+0400 phase correct} is given to show the influence of all possible DMs on the complex vector sum. 
The red solid line in Figure \ref{fig:J1905+0400 phase correct} shows that when the raw data contains a pulsar signal, matching with the correct DM values can maximize the modulus of vector sum at pulsar frequency.
However, Figure \ref{fig:J1905+0400 phase correct} contains only two frequencies, which is too few compared with the all possible frequencies of raw data.

\begin{figure}[htb]
  \centering
  \includegraphics[width=10cm, angle=0]{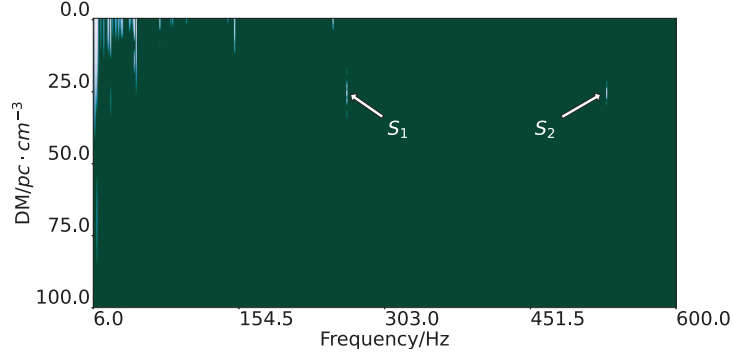}
  \caption{DM-frequency image of PSR J1905+0400. Horizontal axis is frequency while vertical axis is DM. The Brightness of each point is the modulus of complex vector sum with corresponding frequency and DM. In order to facilitate eye observation, the image is enhanced.}
  \label{fig:J1905+0400 DM-frequency image}
\end{figure}

To show the variation of vector sum with DMs at all possible frequencies, we draw Figure \ref{fig:J1905+0400 DM-frequency image}, which is called DM-frequency image later. Essentially, DM-frequency image is the visualization of $X_c(\omega)$ in Equation \eqref{eq:fdd2}, while each vertical line in Figure \ref{fig:J1905+0400 DM-frequency image} can be drawn like a curve in Figure \ref{fig:J1905+0400 phase correct}. 
In addition, the goal of time-domain dedispersion shown in Figure \ref{fig:search process} is obtaining a horizontal line of DM-frequency image. 
Therefore, pulsar phase characteristics presented in this paper can be seen as another perspective of the DM-frequency image.

\begin{figure}[htb]
   \centering
   \includegraphics[width=10cm, angle=0]{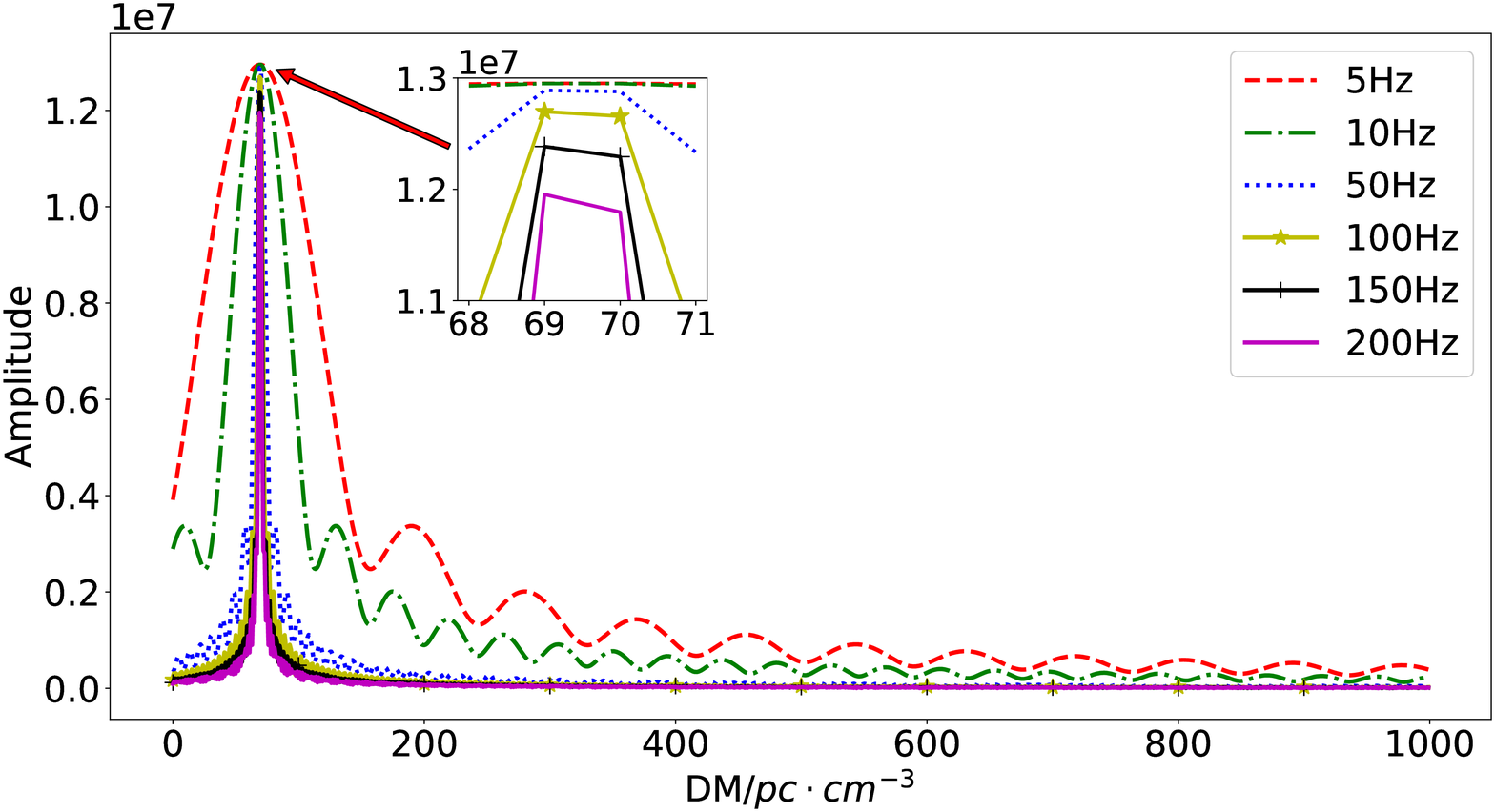}
   \caption{Fake signal of different frequencies with DM 69.55 $\rm {cm}^{-3}\rm{pc}$. Use 0-1000 $\rm {cm}^{-3}\rm{pc}$ as DM with a step size of 1 $\rm {cm}^{-3}\rm{pc}$ to correct the signal, and superimpose the corrected data of each channel to take the modulus.}
   \label{fig:fake signal of different frequencies}
\end{figure}

Since most RFI signals in radio pulsar data are 0 DM signals, the responses of RFIs showed in Figure \ref{fig:J1905+0400 DM-frequency image} are lines starting from 0 $\rm {cm}^{-3}\rm{pc}$, and the closer they approach to 0, the higher the energy they are of. Excluding RFI signals, candidate frequencies can be filtered to several one ($S_1$, $S_2$) in the center of the image, which is PSR J1905+0400 and its harmonic. 
Comparing $S_1$ and $S_2$ in the image, which has the same DM, but with a difference in the length of the signal, $S_2$ with higher frequency is much shorter than $S_1$. To make this phenomenon more clear, we draw a simulation image in Figure \ref{fig:fake signal of different frequencies}, and simulate phase correction curves for pulsar signals with the same DM but different frequencies.

Figure \ref{fig:fake signal of different frequencies} shows that the simulated signal  shows not only a change in peak width as the frequency changes, but also a ``sinc like'' phenomenon when the signal changes with DM. 
For the change of the width on the peaks, it can be explained from the phase correction equation. With the decrease of frequency $\omega _i$, $d_n$'s influence on the rotation factor $ e^{-j\omega _i d_n}$ reduces. 
In other words, under the same DM, the rotation angle of the rotation factor becomes smaller as the candidate frequency decreases, so the higher the signal frequency, the narrower the peak. For a similar reason, the ``sinc like'' function pattern appears in the curve. For channels close to channel 1, $d_n$ is close to 0, a slight change in rotation factor, different channels have different rotation angles. When the phase changes with DM, some of the channels point in the opposite direction to other channels (as the solid line in Figure \ref{fig:J1905+0400 phase align}(a)). The vector sum will appear in the valley area of the curve when these channels account for a large proportion, and it will appear in the peak area with the increase of the similar channels. To those DMs larger than 69.55 $\rm {cm}^{-3}\rm{pc}$, the phase angles of different channels become chaotic and the curve becomes flat with the increase of DM value.

\begin{figure}[htb]
	\begin{minipage}[t]{0.498\linewidth}
		\centering
		\includegraphics[width=74mm]{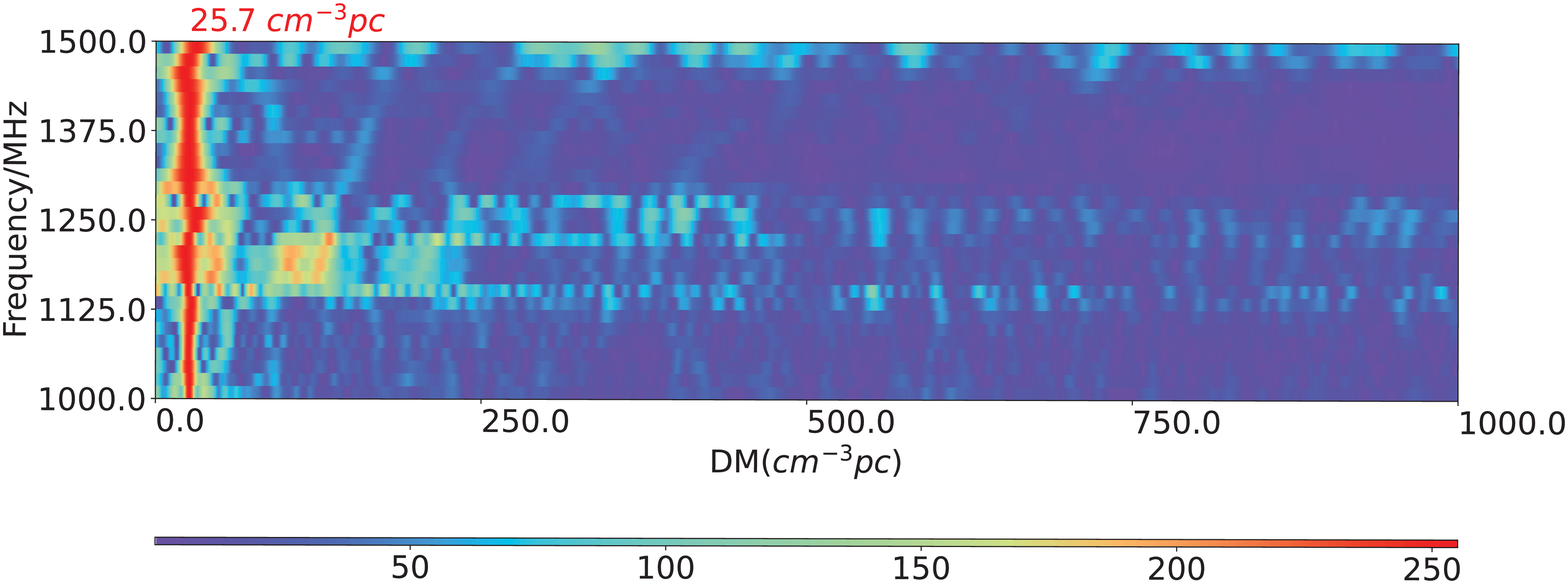}
		\centerline{(a) Multi-channel phase alignment of PSR J1905+0400 ($S_1$)}
	\end{minipage}  
	\begin{minipage}[t]{0.498\linewidth}
		\centering
		\includegraphics[width=74mm]{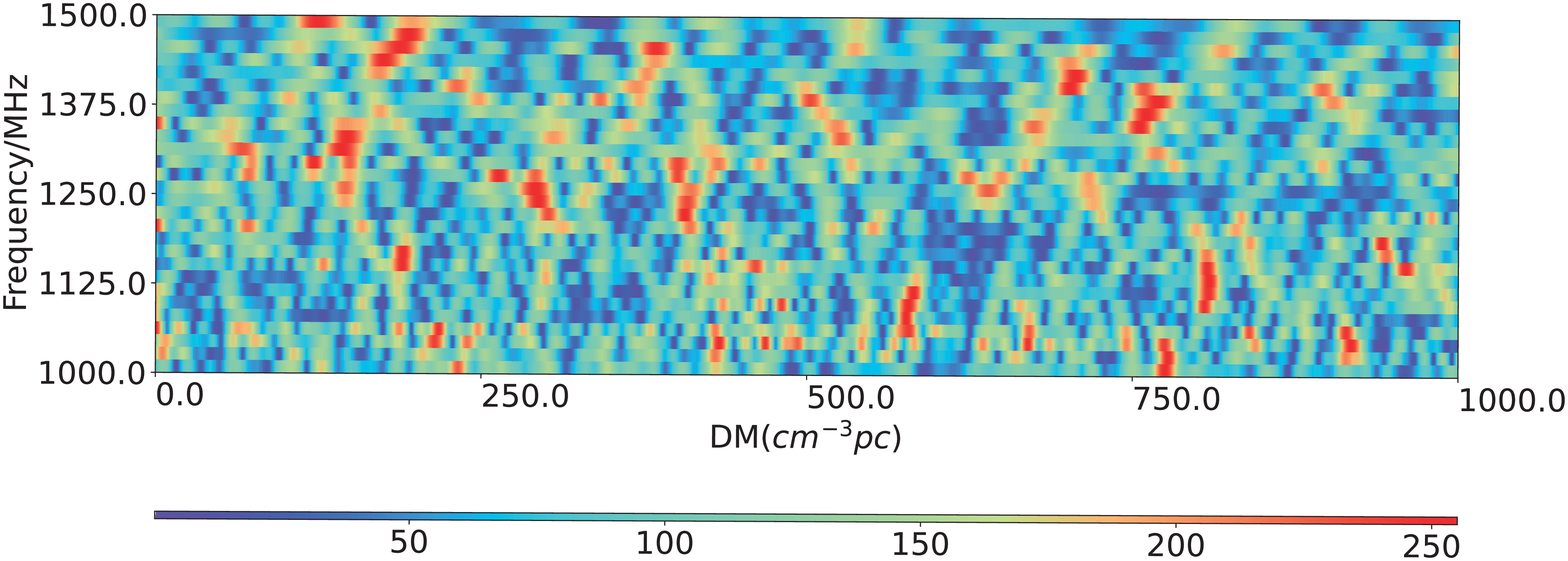}
		\centerline{(b) Multi-channel phase alignment of 200 Hz}
	\end{minipage}
	\caption{Multi-channel phase alignment of PSR J1905+0400 ($S_1$) and 200 Hz. each row in the figure is the vector sum of a channel group, which contains 128 channels and with a channel step of 64, that is group1 contains channel 1-128 and group2 contains channel 65-192, and so on.}
	\label{fig:J1905+0400 S1}
\end{figure}

Based on DM-frequency image, the search for candidate frequencies can be performed, as such an image can be used to check if a signal with dispersion effects exists. 
But when it comes to searches in observed data, just relying on DM-frequency image is not enough, just like searches can not merely rely on pulse profiles for pulsar identification. 
If strong RFIs or noises exist in some channels, there will still be bright spot in DM-frequency image at the RFI frequency.
To distinguish the bright spot brought by RFI and pulsar, we use the broadband characteristic of pulsar signal, that is pulsar signals exist in most channels. 
To reflect broadband characteristic in frequency domain, we divide channels into groups and perform a group by group accumulation at candidate frequency. Then, the broadband characteristic can be reflected with a 2D image like Figure \ref{fig:J1905+0400 S1}, which we call multi-channel phase alignment.

With the multi-channel phase alignment image, candidate frequencies in DM-frequency image like $S_1$ in Figure \ref{fig:J1905+0400 DM-frequency image} can be further judged. The broadband characteristic of $S_1$ has been presented in Figure \ref{fig:J1905+0400 S1} (a), which is almost a straight line throughout the observed frequencies. For the noise like frequency 200Hz we chosen, broadband characteristic does not appear, which is an effective differentiation. In addition, to those RFI signals with a DM value of 0 (as Figure \ref{fig:J1905+0400 phase align} (c)), such image is similar to moving the vertical line to 0 $\rm {cm}^{-3}\rm{pc}$ in Figure \ref{fig:J1905+0400 S1} (a).
To the difference between broadband and narrowband for signals is whether the strongest positions of each channel group are close to form a straight line.

\subsection{Data Filtering and Search Structure Design}
\label{sect: SEARCH WITH PHASE CHARACTERISTIC}

As candidates frequencies can be filtered with DM-frequency image and further judged with multi-channel phase alignment image, a complete search framework can be built. 
However, to draw a DM-frequency image like Figure \ref{fig:J1905+0400 DM-frequency image}, we need traverse all frequencies, just like traverse all DMs in time-domain, which means a huge amount of computation. 
Noting that Figure \ref{fig:J1905+0400 DM-frequency image} only contains a few candidate frequencies with energy anomalies, while most of the data are abandoned in the subsequent calculations, which means phase corrections of those abandoned data are not necessary, and phase corrections can be performed for a few high energy frequencies rather than all frequencies.

To obtain high energy frequencies, through summarizing the experience of data processing, we propose an empirical method. 
The method calculates the sum of each channel's power spectrum, which is the radius of the outer black circle in Figure \ref{fig:J1905+0400 phase align}, and filters top ranked frequencies. 
According to the processing results of PMPS (Parkes Multibeam Pulsar Survey) data and other data, the filtering ratio we recommend is 1\%.
Obviously, it is impossible that all of the top 1\% are candidate frequencies, so further filtering is performed on the top 1\% frequencies after phase corrections. 
As shown in Figure \ref{fig:J1905+0400 phase align}, the correct DM and frequency align complex vectors of each channel, and makes the complex vector sum closer to the outer black circle than other DMs or frequencies.
Therefore, we use the ratio of the length of complex vector sum to the sum of each channel's power spectrum as a further filter parameter.

\begin{figure}[htb]
  \centering
  \includegraphics[width=14.5cm, angle=0]{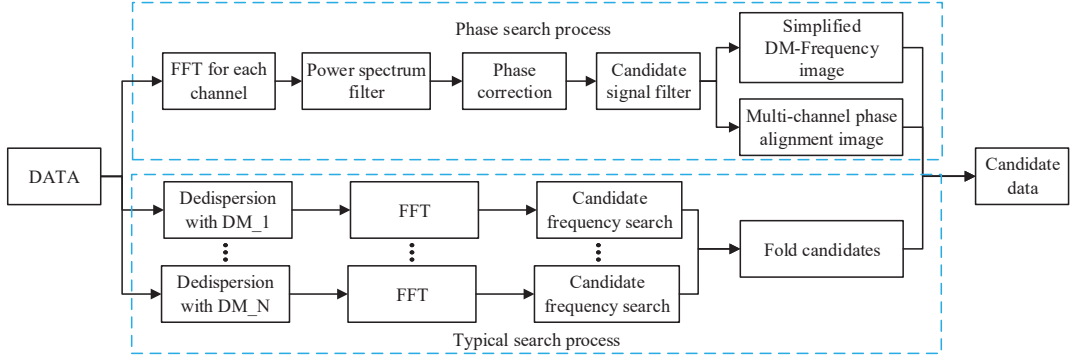}
  \caption{Phase search process compare with typical search process.}
  \label{fig:search compare}
\end{figure}

After the two-step filtering, we obtain the multi-channel phase alignment images of candidate frequencies and simplified DM-Frequency image for manual confirmation. The whole search processing flow is shown in Figure \ref{fig:search compare}, in which the search structure is compared with the typical search method. Our processing code\footnote{\url{https://github.com/hqc-info/dmselect}} is already placed on GitHub.

\section{RESULTS}
\label{sect: results}
\subsection{Test Data}
\label{sect:data test}
In this part, a verification of the phase characteristic based searching method is performed with a group of test data, which contains two globular clusters with short observation time\footnote{\url{https://fast.bao.ac.cn/cms/article/172/}} and two GPPS data files. 
Each GPPS data can survey a cover of a sky area of 0.1575 square degrees and contains 76 adjacent observation points.
The observation information of all test data is shown in Table \ref{tab:obs file info}.
In these test data, several pulsars have been involved, which are listed in Table \ref{tab:Pulsar info}.

\begin{table}[htb]
\begin{center}
\caption[]{Observation Info}
\label{tab:obs file info}
 \begin{tabular}{cccccc}
  \hline\noalign{\smallskip}
Obs File &  Central Freq &Band &Channel &Sample Time &Obs Time\\
  \hline\noalign{\smallskip}
M15 (J2129+1210/B2127+11)  & 1250MHz & 500MHz   &4096  &98.304$\mu$s  & 257.698s \\ 
M71 (J1953+1846)  & 1250MHz & 500MHz   &4096  &49.152$\mu$s  & 180.388s \\
G38.02-1.36 (GPPS) & 1250MHz & 500MHz   &2048  &49.152$\mu$s  & 300s \\
G38.17+3.13 (GPPS) & 1250MHz & 500MHz   &2048  &49.152$\mu$s  & 300s \\
  \noalign{\smallskip}\hline
\end{tabular}
\end{center}
\end{table}

\subsection{Time Performance}
Firstly, we discuss the time performance of our method, and use PRESTO as a comparison.
Table \ref{tab:obs file calculate time} shows the processing time cost on different test data.
In Table \ref{tab:obs file calculate time}, \emph{PRESTO search} refers to the time spent in dedispersion, FFT and acceleration search, while the dedispersion plan is given in \ref{tab:Dedispersion Scheme for Test Data}. 
With the same dedispersion plan, we performed phase characteristic search with CPU and GPU respectively, and the time consumption shown in Table \ref{tab:obs file calculate time} refers to the time spent in FFT, power spectrum filter, phase correction and candidate signal filter shown in Figure \ref{fig:search compare}.
All tests are handled by the same NH55AF laptop\footnote{CPU:3900X, GPU:RTX2070m, RAM:32G, SSD:PM981 512G}, while both \emph{PRESTO search} and \emph{phase characteristic search with CPU} in Table \ref{tab:obs file calculate time} are calculated with one thread.  
Since our method has not done any special processing for binary pulsars, the acceleration search in PRESTO is turned off and the related parameter ZMAX is set to 0. 
From Table \ref{tab:obs file calculate time}, our method shows a speedup of about 4 times than PRESTO with CPU, which can be further accelerated about 8 times with GPU. 

\begin{table}[htb]
	\begin{center}
		\caption[]{Observation File Calculate Time}
		\label{tab:obs file calculate time}
		\begin{tabular}{ccccc}
			\hline\noalign{\smallskip}
			Obs File & DM Range &PRESTO&Phase Characteristic & Phase Characteristic\\ 
			&&Search Time & Search with CPU& Search with GPU\\
			\hline\noalign{\smallskip}
			M15 (J2129+1210/B2127+11) & 0-1000$\rm {cm}^{-3}\rm{pc}$ &1207.04s & 330.83s   &42.14s \\ 
			M71 (J1953+1846) & 0-1000$\rm {cm}^{-3}\rm{pc}$ &1779.94s & 477.91s   &62.21s \\
			GPPS file\tablefootnote{One of observation point, randomly selected.}  & 0-1000$\rm {cm}^{-3}\rm{pc}$ &2358.18s & 494.35s  &46.91s \\
			\noalign{\smallskip}\hline
		\end{tabular}
	\end{center}
\end{table}

\subsection{Sensitivity Comparison}

\begin{figure}[htb]
\begin{minipage}[t]{0.498\linewidth}
  \centering
   \includegraphics[width=74mm]{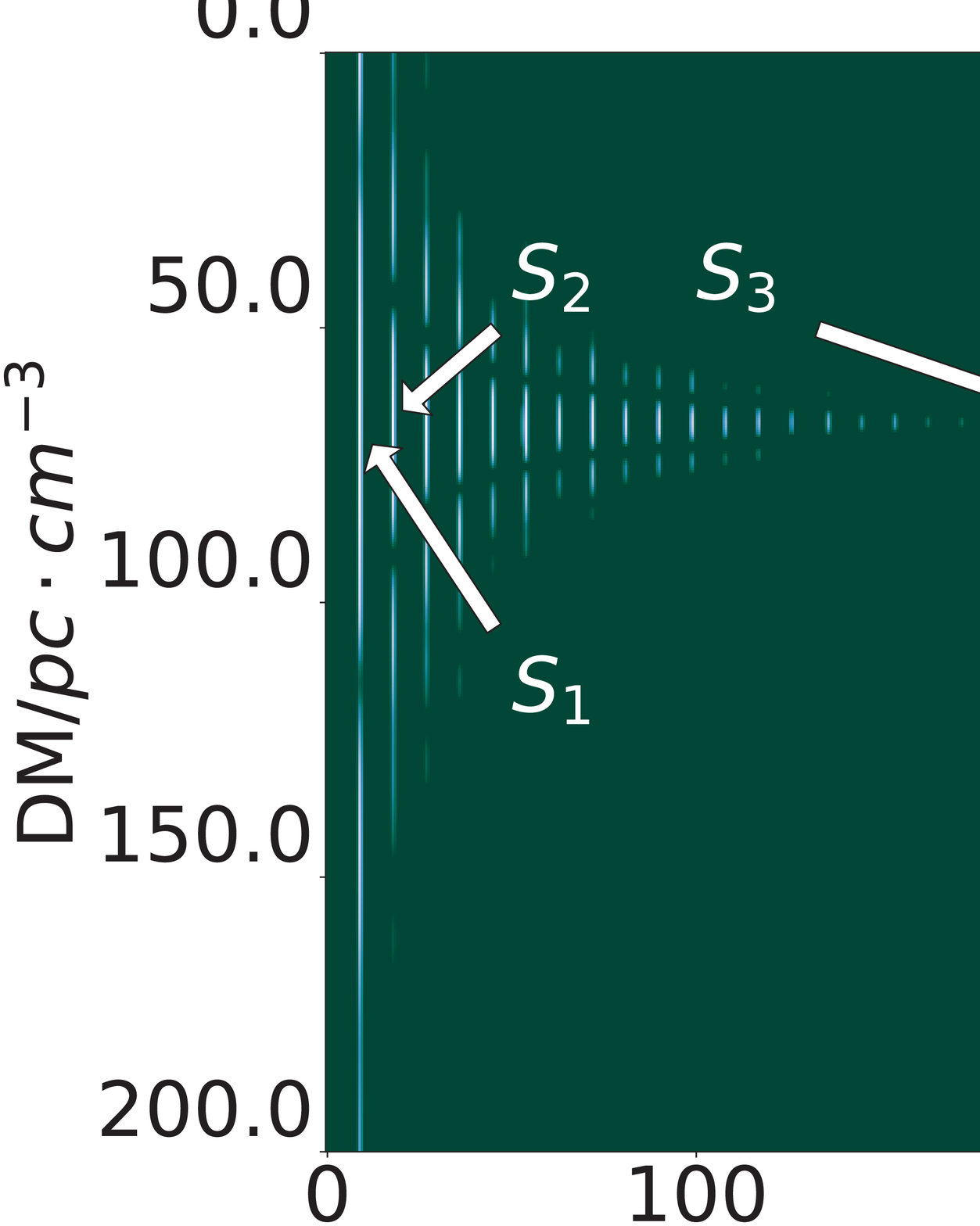}
    \centerline{(a) Simplified DM-Frequency image of M15}
  \end{minipage}  
\begin{minipage}[t]{0.498\linewidth}
  \centering
   \includegraphics[width=74mm]{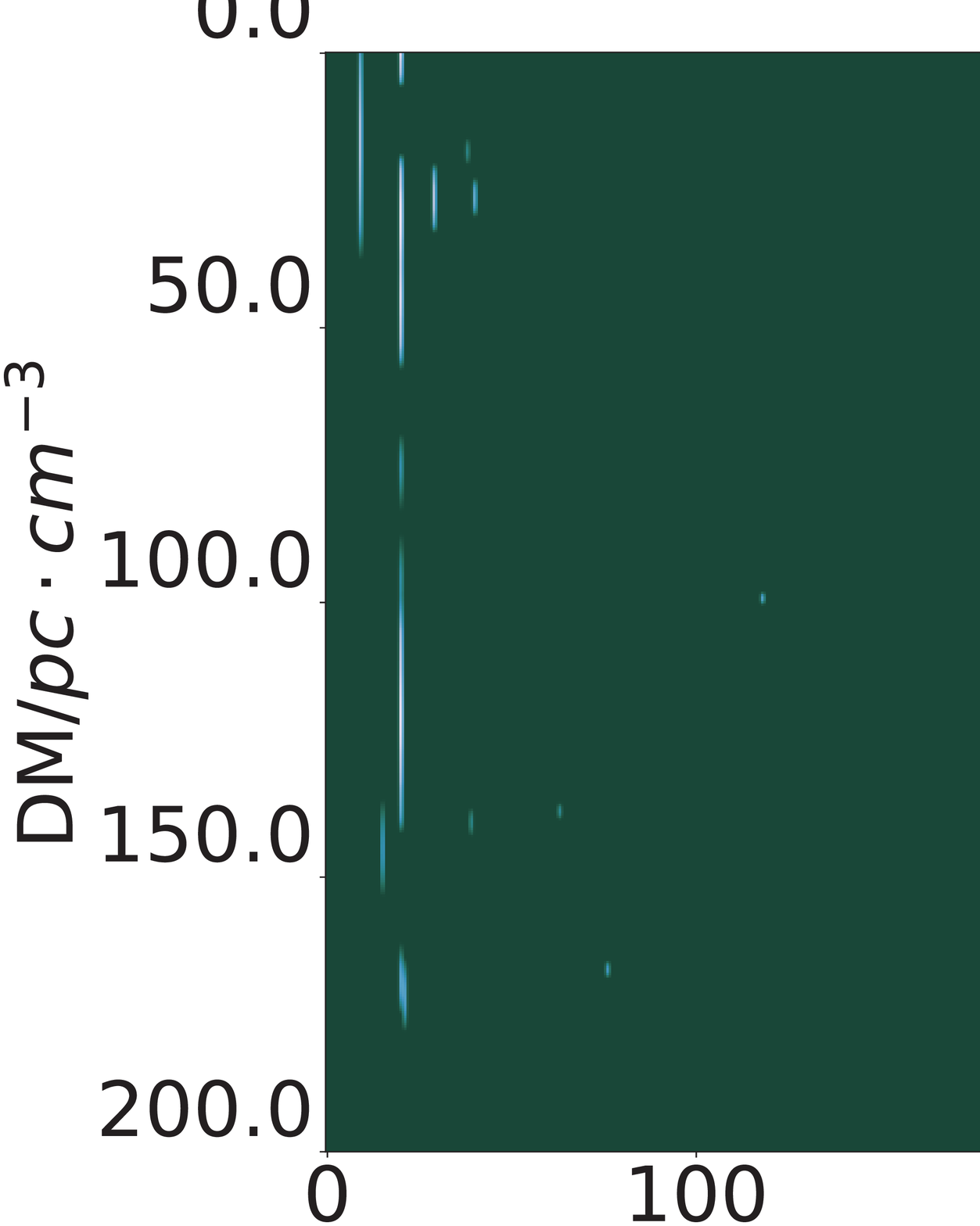}
    \centerline{(b) Simplified DM-Frequency image of M71}
  \end{minipage}
   \caption{M15 and M71, simplified DM-Frequency image, Removed some high-energy signal with low frequency (below 6 $\rm{Hz}$). }
  \label{fig:M15 and M71 data with 2D}
\end{figure}

\begin{figure}[htb]
	\centering
	\includegraphics[width=80mm]{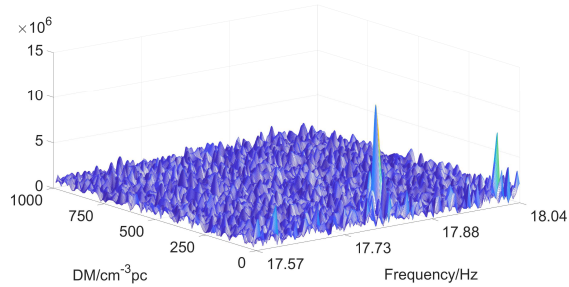}
	\caption{PSR J2129+1210B, phase correction of the frequency points near the signal. Using 0-1000 $\rm {cm}^{-3}\rm{pc}$ to correct the frequency points of different frequencies, the z-axis is the modulus for each channel's sum after the phase of the frequency points is corrected.}
	\label{fig:3D multi pulsar phase correction}
\end{figure}

After the search with our method, hundreds of possible signals will be obtained, which can be drawn as a simplified DM-Frequency image like Figure \ref{fig:M15 and M71 data with 2D} for fast manual search. 
However, in Figure \ref{fig:M15 and M71 data with 2D}(a), M15's DM-Frequency image contains PSR J2129+1210A (9.03 $\rm{Hz}$) and its numerous harmonics. 
These harmonics even masked other signals, making it hard to distinguish $S_2$ (PSR J2129+1210B, 17.81$\rm{Hz}$) from J2129+1210A’s harmonic $S_1$ in the image. 
To distinguish $S_2$ from $S_1$, we can draw a 3D image like Figure \ref{fig:3D multi pulsar phase correction}, which shows that the harmonic of $S_1$ covers other signal. 

Figure \ref{fig:3D multi pulsar phase correction} can be regarded as the local amplification of Figure \ref{fig:M15 and M71 data with 2D}(a) around $S_2$, but it is not necessary to draw a 3D diagram like Figure \ref{fig:3D multi pulsar phase correction}.
In practical use, we suggest using the simplified DM-Frequency image as a tool to quickly determine whether there is any pulsar candidate existing, then separating the possible candidates with the multi-channel phase alignment images manually.
Here using M15 as an example. It can be preliminarily confirmed that one or more pulsars exist in M15 from Figure \ref{fig:M15 and M71 data with 2D}(a). 
Then, we check all the multi-channel phase alignment images generated, eliminate the influence of harmonics and RFI manually, and obtain pulsar candidate signals ($S_1$, $S_2$, $S_3$) with M15 database, which is actually the PSR J2129+1210A, B and D.

\begin{figure}[htb]
	\begin{minipage}[t]{0.6\linewidth}
		\centering
		\includegraphics[width=76mm]{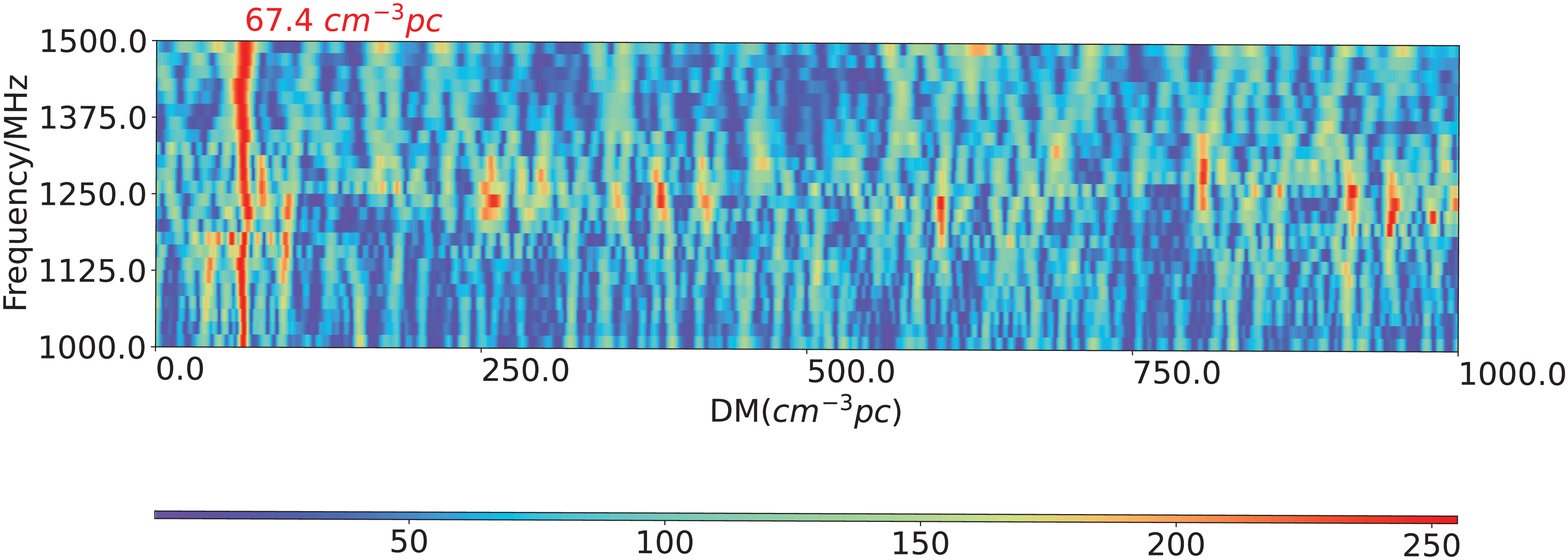}
		\centerline{(a) Multi-channel phase alignment of PSR J2129+1210D (M15D)}
	\end{minipage}
	\begin{minipage}[t]{0.39\textwidth}
		\centering
		\includegraphics[width=55mm]{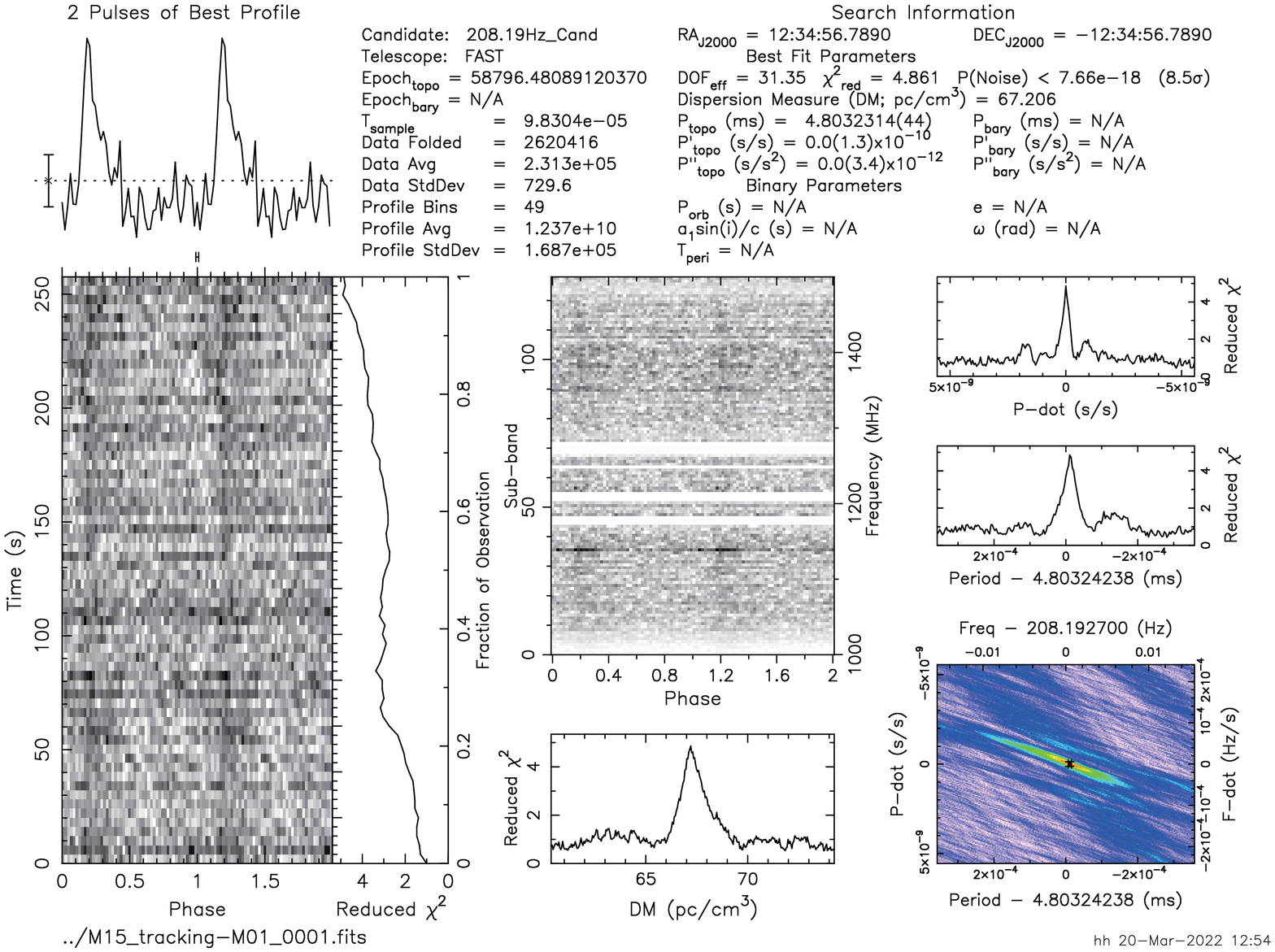}
		\centerline{(b) PSR J2129+1210D fold by PRESTO}
	\end{minipage}
	\caption{Multi-channel phase alignment plot and PRESTO fold plot.}
	\label{fig:M15D data compare}
\end{figure}

According to the search results of PRESTO, the signal-to-noise ratio of PSR J2129+1210D ($S_3$ in \ref{fig:M15 and M71 data with 2D}(a)) is even weaker than the 37th harmonic of PSR J2129+1210A, but it is still remarkable in Figure \ref{fig:M15D data compare} (a).
As a comparison, we give PRESTO's folding result of PSR J2129+1210D in Figure \ref{fig:M15D data compare} (b), in which the sub-bands folding diagram is similar to Figure \ref{fig:M15D data compare} (a).
These two diagrams have a great similarity in the expression of broadband characteristics, except that one uses time domain for presentation and the other uses frequency domain, while the contrast of frequency domain diagram is stronger.

The search results of test methods and the known pulsars are placed in Table \ref{tab:known pulsars}. Based on the data of the FAST Globular Cluster Pulsar Survey (\citealt{pan2021fast}), there should be 14 pulsars in M15 and M71: PSR J2129+1210A-I, PSR J1953+1846A-E. Given that we used data with a much shorter observation time than them, the number of detectable pulsars in test data should be less than 14.
Compared with the PRESTO processing results, the number of pulsars found with the phase characteristic search method is almost consistent with the detectable pulsars in test data.

\begin{table}[htb]
\begin{center}
\caption[]{Search Results}
\label{tab:known pulsars}
 \begin{tabular}{cccc}
  \hline\noalign{\smallskip}
Obs File & The Known Pulsars on& PRESTO Search& Pulsars Found by Our Phase \\
&Observation Point Position\tablefootnote{Data from ATNF website \url{https://www.atnf.csiro.au/research/pulsar/psrcat/} and GPPS website \url{http://zmtt.bao.ac.cn/GPPS/GPPSnewPSR.html}} & Results on Obs File &Characteristic Method on Obs File\\
  \hline\noalign{\smallskip}
M15 (J2129+1210/B2127+11)& PSR J2129+1210A-I & PSR J2129+1210A, B, D& PSR J2129+1210A, B, D\\ 
M71 (J1953+1846) & PSR J1953+1846A-E &PSR J1953+1846A &PSR J1953+1846A\\
G38.17+3.13 (GPPS)& PSR J1848+0604& PSR J1848+0604& PSR J1848+0604\\
G38.02-1.36 (GPPS)& PSR J1905+0400 & PSR J1905+0400& PSR J1905+0400 \\
&PSR J1906+0352g\tablefootnote{name suffix with "g" indicates the temporary nature}&PSR J1906+0352g&\\
  \noalign{\smallskip}\hline
\end{tabular}
\end{center}
\end{table}

\begin{figure}[htb]
 \begin{minipage}[t]{0.495\linewidth}
  \centering
  \includegraphics[width=74mm]{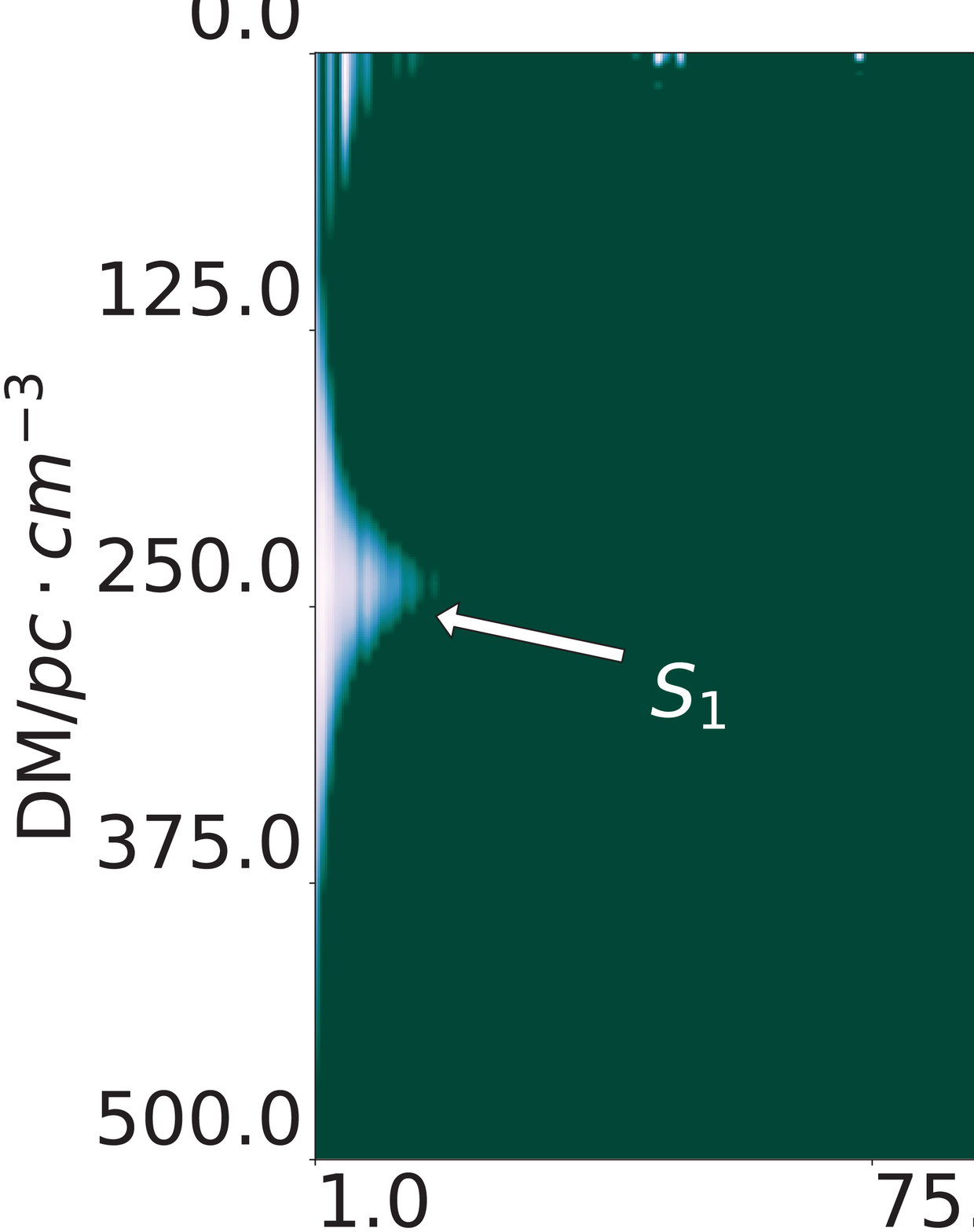}
    \centerline{(a) PSR J1848+0604 DM-Frequency image}
  \end{minipage}
  \begin{minipage}[t]{0.495\textwidth}
  \centering
  \includegraphics[width=74mm]{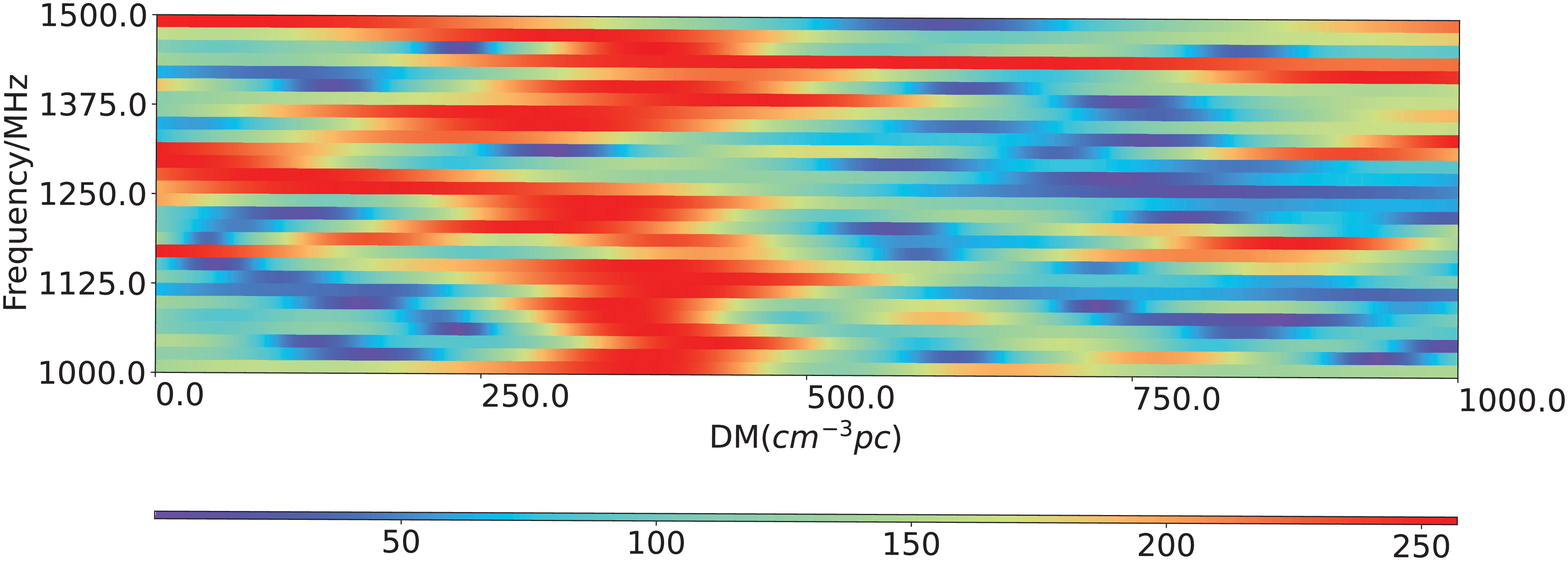}
  \centerline{(b) PSR J1906+0352g's 4th harmonic}
  \end{minipage}
\caption{(a) is PSR J1848+0604 DM-Frequency image. (b) is the Multi-channel phase alignment image of PSR J1906+0352g's 4th harmonic.}
  \label{fig:GPPS data result}
 \end{figure}

For the two GPPS test files, after querying and comparing on the ATNF website, two known pulsars can be found: PSR J1848+0604 and PSR J1905+0400 (in Section \ref{sect:Phase Alignment}). However, the PSR J1906+0352g found by \cite{han2021fast} cannot be sought out by our method, while only the harmonic of PSR J1906+0352g can be found. In Figure \ref{fig:GPPS data result}, (a), (b) correspond to PSR J1848+0604 and PSR J1906+0352g's 4th harmonic respectively.
PSR J1848+0604 has a low frequency, the resolution of the drawn image is not enough to distinguish its frequency. But the higher energy makes it easier to distinguish from the data. To the signal of PSR J1906+0352g, it is much weaker than PSR J1848+0604, due to the influence of noise, only some harmonics can be found. For such missing signal detection, it reminds us of the necessity of harmonic processing for phase characteristics which have a great help for weak signal detection.

\section{DISCUSSION AND CONCLUSION}
\label{sect:conclusion}
We have presented a method making use of phase characteristics for pulsar search and tested its feasibility by processing the data of M15, M71, and so on. The pulsar signals are searched and confirmed in the light of their characteristics. Such method can be used to achieve faster search, and the phase characteristic can also be integrated into the traditional search structure with FDD, only needs to save the intermediate processing data during FDD processing. In the candidate signal judgment step of the pulsar search, DM-Frequency images can quickly assist researchers to establish the relationship between the frequency of the data and the DM, especially for pulsars with many harmonics. Based on the two-step filtering, only a few frequencies need to be processed to verify the broadband characteristic. Therefore, the processing scale required by our method is much smaller than PRESTO, so the calculation time is also reduced.

For our method, there are still some problems that need to be solved, the known problems are as follows: (1) The screening in the existing search process is carried out with empirical parameters, how to build an accurate mathematical model to complete the screening. And, in candidate filtering, there may be a more suitable probability model, how to optimize the existing model and add it to the search process could be a problem. In addition, the further processing of harmonics is also needed to add. (2) The phase characteristics still depend on the calculation results of FFT. This means it is difficult to find those low-frequency signals masked by red noise, which is also faced by other search methods (\citealt{van2017framework}). (3) Since there is no additional algorithm adjustment for binary pulsars, it is hard to say that the phase characteristic which is obviously suitable for some bright data such PSR J1953+1846A can also fit for faint binary pulsars data. (4) For some very short orbit period binary pulsars or binary pulsars of long observation time data, they will produce a palpable frequency shift effect when they rotate (\citealt{ransom2003new}). The pulse phase of the received data after Fourier transform can not reflect the information of the signal, which also needs to be concerned. However, to one’s delight, it is at least clear that what kind of data can be processed by our method.

\acknowledgements
This work is supported by the National Natural Science Foundation of China (No. 12203039 and No. 11873083). This work made use of the data from FAST (Five-hundred-meter Aperture Spherical radio Telescope). FAST is a Chinese national mega-science facility, operated by National Astronomical Observatories, Chinese Academy of Sciences. This work is also supported by National SKA Program of China (No. 2020SKA0120100),  National Nature Science Foundation of China (No. 12173053 and No. 12041303), the Youth Innovation Promotion Association of CAS (id.~2018075), the CAS "Light of West China" Program,  the Specialized Research Fund for State Key Laboratories, and the Science and Technology Program of Guizhou Province ([2021]4001).

\appendix                  
\section{Pulsars Info}

\begin{table}[htb]
\begin{center}
\caption[]{Pulsars Info Involved in This Paper}
\label{tab:Pulsar info}
 \begin{tabular}{ccccc}
  \hline\noalign{\smallskip}
PSR & F0 (Hz)& DM ($\rm {cm}^{-3}\rm{pc}$)& References &Notes\\
  \hline\noalign{\smallskip}
 J2129+1210A (M15A/B2127+11A)  & 9.03 & 67.31& \citealt{anderson1993study}\\ 
 J2129+1210B (M15B/B2127+11B)  & 17.81 & 67.69& \citealt{anderson1993study}\\
 J2129+1210D (M15D/B2127+11D)  & 208.21& 67.3& \citealt{anderson1993study}\\
 J1953+1846A (M71A)  & 204.58& 117& \citealt{hessels20071}\\
 J1848+0604  & 0.4507& 242.7& \citealt{lorimer2005discovery}\\
 J1905+0400  & 264.24& 25.69& \citealt{gonzalez2011high}\\
 J1906+0352g & 3.50 & 372&\citealt{han2021fast}& Pulsar was discovered after\\
 &&&& the publication of the paper\\
  \noalign{\smallskip}\hline
\end{tabular}
\end{center}
\end{table}

\begin{table}[htb]
\begin{center}
\caption[]{Details on Dedispersion Plan for Test Data}
\label{tab:Dedispersion Scheme for Test Data}
 \begin{tabular}{ccccccc}
  \hline\noalign{\smallskip}
DATA & DM Start& DM End& DM Step& Down Sample & Number of DMs& Notes\\
&($\rm {cm}^{-3}\rm{pc}$)&($\rm {cm}^{-3}\rm{pc}$)&($\rm {cm}^{-3}\rm{pc}$)&&\\
  \hline\noalign{\smallskip}
M15& 0.0& 1005.0 & 0.3& 4& 3350& prepsubband function with 256 nsub \\
M71& 0.0& 1005.0 & 0.3& 4& 3350& prepsubband function with 256 nsub\\
GPPS File& 0.0& 1005.0 & 0.3& 4& 3350& prepsubband function with 256 nsub\\
  \noalign{\smallskip}\hline
\end{tabular}
\end{center}
\end{table}

\bibliographystyle{raa}
\bibliography{bibtex}

\label{lastpage}
\end{document}